\documentstyle[aps,multicol,epsf]{revtex}
\input epsf
\draft
\title{Comparison of Recoil-Induced Resonances (RIR) and Collective Atomic Recoil
Laser (CARL)}
\author{P. R. Berman 
\\ {\em Physics Department, University of Michigan, Ann Arbor, Michigan 48109-1120} }
\date{\today }
\begin{document}
\maketitle
\begin{abstract}
The theories of recoil-induced resonances (RIR) [J. Guo, P. R. Berman, B.
Dubetsky and G. Grynberg, Phys. Rev. A {\bf 46}, 1426 (1992)] and the
collective atomic recoil laser (CARL) [ R. Bonifacio and L. De Salvo, Nucl.
Instrum. Methods A {\bf 341}, 360 (1994)] are compared. Both theories can be
used to derive expressions for the gain experienced by a probe field
interacting with an ensemble of two-level atoms that are simultaneously
driven by a pump field. It is shown that the RIR and CARL formalisms are
equivalent. Differences between the RIR and CARL arise because the theories
are typically applied for different ranges of the parameters appearing in
the theory. The RIR limit considered in this paper is $qP_{0}/M\omega
_{q}\gg 1$, while the CARL limit is $qP_{0}/M\omega _{q}\lesssim 1$, where $%
q $ is the magnitude of the difference of the wave vectors of the pump and
probe fields, $P_{0}$ is the width of the atomic momentum distribution and $%
\omega _{q}$ is a recoil frequency. The probe gain for a probe-pump detuning
equal to zero is analyzed in some detail, in order to understand how the
gain arises in a system which, at first glance, might appear to have
vanishing gain. Moreover, it is shown that the calculations, carried out in
perturbation theory have a range of applicability beyond the recoil problem.
Experimental possibilities for observing CARL are discussed.
\end{abstract}

\section{Introduction}

In recent years, there has been increased interest in spectral features that
can attributed directly to the recoil atoms undergo on the absorption,
emission or scattering of radiation. Among topics that have been discussed
that fall into this category are Recoil-Induced Resonances (RIR) \cite
{berman,guo,pumpprobe,faraday,grynberg,gryntrans,metcalf,japanese,raizen}
and the Collective Atomic Recoil Laser (CARL) \cite
{bon,verkerk,propagation,meystre}. These processes appear to have much in
common, although they are described quite differently. It is the purpose of
this article to compare the RIR and CARL and to demonstrate that the RIR and
CARL {\em formalisms} are equivalent. The reason that this equivalence is
not readily apparent is two-fold. Firstly, the RIR have been discussed using
a density matrix approach in the Schr\"{o}dinger picture, whereas CARL has
been discussed using an operator approach in the Heisenberg picture.
Secondly, RIR and CARL have generally been examined for different ranges of
the various parameters characterizing the atom-field interactions. In order
to facilitate the discussion, it is useful to review briefly the RIR and
CARL.

Since both the RIR and CARL relate to probe field absorption or gain in the
presence of a pump field, it is perhaps best to recall the features of the
probe absorption spectrum, {\em neglecting all effects associated with recoil%
}. Consider an ensemble of two-level atoms interacting with both a pump and
probe field. The probe absorption or gain is monitored as a function the
probe-pump detuning $\delta $ for various pump field strengths,
characterized by the pump field Rabi frequency $\chi _{1}$. It is assumed
that all collisional effects can be neglected and that the two-level atomic
system is {\em closed} in the sense that the sum of ground and excited state
populations is conserved for {\em each} velocity subclass of atoms.
Moreover, it is assumed that the pump field detuning $\Delta $ from atomic
resonance is much larger than any decay rates, Rabi frequencies, or Doppler
shifts in the problem. In this limit, the probe absorption spectrum consists
of three features, centered near $\delta =0,\pm \Delta $ \cite{mollow}.

Of primary concern here is the structure centered near $\delta =0.$ In the
absence of recoil, this line feature has a dispersion-like shape, is
proportional to $\chi _{1}^{4}$, and has a width of order of the excited
state decay rate \cite{doppler}. Its origin can be traced to an interference
effect involving processes in which {\em two} photons are emitted into
previously unoccupied vacuum mode states \cite{grynpumpprobe}. As such, it
is linked to spontaneous emission, rather than a conservative exchange of
energy between the pump and probe fields. The fact that the two-level system
is ''closed'' plays a critical role here. Had the system been ''open,'' the
amplitude of this component would vary as $\chi _{1}^{2}$ rather than $\chi
_{1}^{4}$ and its width could be determined by some effective ground state
decay rate (e.g. inverse transit time) rather than the excited state decay
rate \cite{galina}.

The RIR refer to a class of spectroscopic features in nonlinear spectroscopy
that can be attributed to a recoil-induced ''opening'' of an otherwise
closed, two-level system \cite{berman}. In the presence of recoil, the
atomic velocity is changed on the absorption or emission of radiation. As a
consequence the sum of ground and excited state populations for specific
velocity classes is no longer conserved. In the limit of large detuning $%
\Delta $, the RIR can be interpreted as arising from Raman transitions
between different center-of-mass states \cite{berman}. If the pump and probe
fields have propagation vectors {\bf k}$_{1}$ and {\bf k}$_{2}$,
respectively, then probe absorption occurs on a Raman transition between
center-of-mass momentum states $\left| {\bf P}\right\rangle $ and $\left| 
{\bf P-}\hbar {\bf q}\right\rangle $ and gain between states $\left| {\bf P}%
\right\rangle $ and $\left| {\bf P+}\hbar {\bf q}\right\rangle ,$ where $%
{\bf q=}${\bf k}$_{1}-${\bf k}$_{2}$. Depending on the detuning $\delta $
between the probe and pump fields, one of these processes is favored. For $%
\delta <0$, there is probe gain, for $\delta >0,$ there is probe absorption,
and for $\delta =0,$ the nonlinear probe absorption and gain vanish. In
general, the RIR have been formulated as a {\em stationary} process in which
the field achieves a steady-state value at each position ${\bf R}$ - gain
occurs as the probe field propagates through the active medium. Steady state
is achieved by assuming that there is some effective ground state decay rate
or, alternatively, by assuming that the Doppler width associated with the
Raman transitions, $qP_{0}/M,$ where $M$ is an atomic mass and $P_{0}$ is
the width of the momentum distribution, is larger than the recoil frequency, 
$\omega _{q}=\hbar q^{2}/2M$ \cite{gryntrans,japanese,boris}. The gain is
normally expressed as $dE_{2}/dZ=\alpha E_{2}$, where $E_{2}$ is the probe
field's amplitude and $\alpha $ is a constant proportional to the atomic
density. The point to note here is that the derivative of $E_{2}$ depends 
{\em locally} on $E_{2}.$

In contrast to the RIR, CARL has been formulated as a {\em transient}
problem. It is assumed that the pump and probe fields are modes of an
optical cavity in which the atoms are located. As a result of the atom-field
interaction, the probe field builds up in the cavity as a function of time.
In the limit of large $\Delta ,$ one obtains an equation of the form $%
dE_{2}/dt=\beta \int_{0}^{t}f(t,t^{\prime })E_{2}(t^{\prime })dt^{\prime }$,
where $\beta $ where is a constant proportional to the atomic density and $f$
is a kernel. If $qP_{0}/M<\omega _{q}$, the time derivative of $E_{2}$ can
depend on the past history of the field, rather than locally on the field.
This leads to very different dynamics than those encountered in the RIR. One
finds threshold conditions for probe gain that depend nonlinearly on the
atomic density, as does the gain itself. CARL is a cooperative effect in the
sense that gain occurs only for some minimum atomic density for most values
of the detuning $\delta $. It is possible to have gain for $\delta \geq 0,$
which is not possible for the RIR. On the other hand, if $qP_{0}/M\gg \omega
_{q}$, the equation for $dE_{2}/dt$ becomes local and the gain
characteristics of CARL become identical to those of the RIR. Thus, although
both the RIR and CARL can be formulated for arbitrary ratios of $%
qP_{0}/M\omega _{q}$, one can label the RIR limit as one in which $%
qP_{0}/M\gg \omega _{q}$ and the CARL limit as one in which $%
qP_{0}/M\lesssim \omega _{q}$ \cite{cubeQ}$.$ This is the nomenclature that
is adopted in this article. The Carl limit can be stated as, $2P_{0}/\hbar
q\lesssim 1$, which places it at or near the {\em subrecoil limit} of atom
cooling. It should be noted that Verkerk and Bonifacio \cite{verkerk} and
Bonifacio {\it et al.} \cite{propagation} have shown that the CARL equations
reproduce the RIR results in the limit that $qP_{0}/M\gg \omega _{q}$;
however the approach they followed differs considerably from the one to be
presented herein.

It is important to recognize that it is possible for $dE_{2}/dt$ to depend
locally $E_{2}(t)$, even in the ''CARL limit,'' $qP_{0}/M\lesssim \omega
_{q} $. For example, if one includes ground state decay with rate $\Gamma $,
a local dependence occurs if $\Gamma \gg \omega _{q};$ as a consequence, in
the presence of decay, the RIR limit can be stated as one in which $%
qP_{0}/M\gg \omega _{q}$ {\em or} $\Gamma \gg \omega _{q}$ and the CARL
limit as one in which $qP_{0}/M\lesssim \omega _{q}$ {\em and} $\Gamma
\lesssim \omega _{q}$ \cite{cubeQ}. Moreover, the build up of the probe
field at very early times always varies linearly with the atomic density and
can be viewed as a local process. In this sense, the theory of RIR is always
valid at early times - the specific conditions are given in Sec. III.
However, when $qP_{0}/M\lesssim \omega _{q}$ and $\Gamma \ll \omega _{q}$,
the gain coefficient in CARL depends nonlinearly on the atomic density for
times $t\gtrsim \omega _{q}^{-1}$ \cite{cubeQ}, and the local dependence no
longer holds.

In Sec. II, the basic equations are derived and the RIR and CARL limits of
these equations are obtained in Sec. III. An effective potential for ground
state atoms is written which helps to categorize the RIR and CARL limits.
The case of equal pump and probe frequencies is analyzed in Sec. IV using an
effective five-level atom. When expressed in this form the calculation has a
range of applicability beyond the recoil problem. The results are discussed
in Sec. V. Special emphasis is placed on the distinction between ''atom
bunching'' and ''matter gratings.'' Experimental implications of the results
are also explored.

\section{Basic Equations}

The problem under consideration consists of a pump field and a probe field
interacting with an ensemble of two level atoms . In some applications in
CARL, it may be of interest to use a quantized description of the fields to
follow the build-up of the probe field from noise, but, in the present
discussion, the fields are taken to be classical. The pump field, 
\begin{equation}
{\bf E}_{1}({\bf R},t)=\frac{1}{2}{\bf \epsilon }_{1}\left[ E_{1}({\bf R}%
,t)\,e^{i\left( {\bf k}_{1}\cdot {\bf R}-\Omega _{1}t\right) }+E_{1}^{\ast }(%
{\bf R},t)\,e^{-i\left( {\bf k}_{1}\cdot {\bf R}-\Omega _{1}t\right) }\right]
,  \label{1}
\end{equation}
has polarization {\bf $\epsilon $}$_{1}$, slowly-varying electric field
amplitude $E_{1}({\bf R},t)$, propagation vector ${\bf k}_{1}$ and frequency 
$\Omega _{1}=k_{1}c$, while the probe field, 
\begin{equation}
{\bf E}_{2}({\bf R},t)=\frac{1}{2}{\bf \epsilon }_{2}\left[ E_{2}({\bf R}%
,t)\,e^{i\left( {\bf k}_{2}\cdot {\bf R}-\Omega _{2}t\right) }+E_{2}^{\ast }(%
{\bf R},t)\,e^{-i\left( {\bf k}_{2}\cdot {\bf R}-\Omega _{2}t\right) }\right]
,  \label{2}
\end{equation}
has polarization {\bf $\epsilon $}$_{2}$, slowly-varying electric field
amplitude $E_{2}({\bf R},t)$, propagation vector ${\bf k}_{2}$ and frequency 
$\Omega _{2}=k_{2}c$. As a result of the nonlinear interaction with the
fields, the probe field can be amplified. If the pump field is initially
much more intense than the probe field, as is assumed, pump depletion during
the early stages of probe amplification can be neglected. Since the
calculation in this paper is limited to the early stages of probe
amplification, I take the pump field amplitude to be constant, $E_{1}({\bf R}%
,t)\sim E_{1}.$

The dynamics of probe field amplification depends on the specific atom-field
geometry. One can envision situations in which the probe field amplitude is
constant in time but varies in space, or is constant in space but varies in
time. The first case is the one considered generally in the RIR, in which 
{\em cw} pump and probe fields interact with atoms in a cell or trap. The
probe field amplitude increases in the direction of ${\bf k}_{2}$ as it
propagates through the medium, but is assumed to have evolved to a
stationary state. The second case is the one considered generally in CARL,
in which the fields correspond to field modes of an optical cavity and the
atoms are located in the cavity. For a ring cavity, the probe field
intensity is spatially isotropic, but increases in time \cite{propagation}.
In certain limits (to be noted below), the spatial gain coefficient of the
RIR multiplied by the speed of light coincides with the temporal gain
coefficient of CARL. In other limits, the dynamics of CARL amplification
differs from the spatial build-up of the field in the RIR. To compare the
RIR and CARL, it is convenient to adopt the cavity model and assume that the
probe field amplitude is a function of time only, $E_{2}({\bf R},t)\sim
E_{2}(t).$ All cavity losses are neglected, as is any ground state decay.

Although CARL is referred to as a collective effect since conditions for
CARL gain depend on the atomic density, each atom in the sample, on average,
interacts with the fields in an identical manner. Thus, it is sufficient to
write the Hamiltonian for a single atom interacting with the fields. The
dependence of the field gain on atomic density is included automatically in
the coupled Maxwell-Bloch equations. In the dipole and rotating wave
approximations, the Hamiltonian for our system is 
\begin{equation}
H=\frac{P^{2}}{2M}+\frac{\hbar \omega }{2}\sigma _{z}+\hbar \sum_{\mu =1}^{2}%
\left[ \chi _{\mu }(t)\,e^{i({\bf k}_{\mu }\cdot {\bf R}-\Omega _{\mu
}t)}\sigma _{+}+adjoint\right] ,  \label{3}
\end{equation}
where ${\bf P}$ is the atomic center-of-mass momentum, $\omega $ is the
transition frequency between the ground state $\left| 1\right\rangle $ and
excited state $\left| 2\right\rangle $ of the atom, $\sigma _{z}=(\left|
2\right\rangle \left\langle 2\right| -\left| 1\right\rangle \left\langle
1\right| ),$ $\sigma _{+}=\left| 2\right\rangle \left\langle 1\right| $, 
\begin{equation}
\chi _{\mu }(t)=-\frac{{\bf \wp }\cdot {\bf \epsilon }_{\mu }E_{\mu }(t)}{%
2\hbar }  \label{4}
\end{equation}
is the Rabi frequency of field $\mu $, and {\bf $\wp $}${\bf \equiv }%
\left\langle 2\right| ${\bf $\wp $}$\left| 1\right\rangle $ is a dipole
moment matrix element. Terms related to spontaneous emission are not
included in the Hamiltonian (\ref{3}), for reasons to be discussed below.
The Hamiltonian determines the time evolution of the atom, but the atomic
evolution must be coupled to the field evolution via Maxwell's equations to
arrive at a closed set of equations.

The time evolution of the (complex) probe field amplitude is linked to the
polarization of the medium which, in turn, is determined by the atom-field
interaction. The medium's polarization can be expressed in terms of atomic
density matrix elements as 
\begin{equation}
{\bf P}({\bf R},t)=N\left[ {\bf \wp \,}\rho _{12}({\bf R},t)+{\bf \wp }%
^{\ast }{\bf \,}\rho _{21}({\bf R},t)\right] ,  \label{5}
\end{equation}
where $N$ is the total number of atoms. As a result of the nonlinear
atom-field interaction, the density matrix element $\rho _{21}({\bf R},t)$
can be written quite generally (see below) as 
\begin{equation}
\rho _{21}({\bf R},t)=\tilde{\rho}_{21}(t;1)\,e^{i\left( {\bf k}_{1}\cdot 
{\bf R}-\Omega _{1}t\right) }+\tilde{\rho}_{21}(t;2)\,e^{i\left( {\bf k}%
_{2}\cdot {\bf R}-\Omega _{2}t\right) }+\sum_{n\neq 0,-1}\tilde{\rho}%
_{21}(t;1,n)\,e^{i\left( {\bf k}_{1}\cdot {\bf R}-\Omega _{1}t\right) +in(%
{\bf q}\cdot {\bf R}+\delta t)},  \label{6}
\end{equation}
where 
\begin{equation}
{\bf q}={\bf k}_{1}-{\bf k}_{2};\qquad \delta =\Omega _{2}-\Omega _{1}.
\label{7}
\end{equation}
In the slowly varying amplitude and phase approximation, it follows from
Maxwell's equations and Eqs. (\ref{5}) and (\ref{6}) that the evolution of
the probe field is given by 
\begin{mathletters}
\label{8}
\begin{eqnarray}
\frac{dE_{2}(t)}{dt} &=&\frac{i\Omega _{2}}{\epsilon _{0}}{\bf \epsilon }%
_{2}\cdot {\bf P}_{2}({\bf R},t)\,e^{-i\left( {\bf k}_{2}\cdot {\bf R}%
-\Omega _{2}t\right) }  \label{8a} \\
&=&\frac{iN\Omega _{2}}{\epsilon _{0}}{\bf \epsilon }_{2}\cdot {\bf \wp }%
^{\ast }\tilde{\rho}_{21}(t;2),  \label{8b}
\end{eqnarray}
where ${\bf P}_{2}({\bf R},t)$ is the component of the polarization ${\bf P}%
_{2}({\bf R},t)$ varying as $e^{i\left( {\bf k}_{2}\cdot {\bf R}-\Omega
_{2}t\right) }.$ Combining Eq. (\ref{8b}) with Eq. (\ref{4}) yields 
\end{mathletters}
\begin{equation}
\frac{d\chi _{2}(t)}{dt}=-\frac{iNd^{2}\Omega _{2}}{2\hbar \epsilon _{0}}%
\tilde{\rho}_{21}(t;2),  \label{9}
\end{equation}
where 
\begin{equation}
d\equiv \left| {\bf \wp }\cdot {\bf \epsilon }_{2}\right| .  \label{10}
\end{equation}
An equation for $\tilde{\rho}_{21}(t;2)$ is obtained by solving the
Schr\"{o}dinger equation for the Hamiltonian (\ref{3}).

It is convenient to expand the wave function of the system as 
\begin{equation}
\psi ({\bf R},{\bf r},t)=\sum_{\mu =1,2}A_{\mu }({\bf R},t)\,\psi _{\mu }(%
{\bf r})  \label{11}
\end{equation}
where 
\begin{equation}
A_{\mu }({\bf R},t)=\frac{1}{\left( 2\pi \hbar \right) ^{3/2}}\int d{\bf P}%
A_{\mu }({\bf P},t)\,e^{-i\omega _{\mu }t}e^{i{\bf P\cdot R}/\hbar
}\,e^{-iE_{P}t/\hbar },  \label{12}
\end{equation}
$\omega _{1}=-\omega /2$, $\omega _{2}=\omega /2$, $E_{P}=P^{2}/2M$, and $%
\psi _{\mu }({\bf r})$ is the atomic-state eigenfunction associated with
state $\mu $. Note that the momentum state amplitudes, $A_{\mu }({\bf P},t)$%
, are defined in an interaction representation. The state amplitudes, $%
a_{\mu }({\bf P},t)$, in the ''normal'' representation are related to those
in the interaction representation by 
\begin{equation}
a_{\mu }({\bf P},t)=A_{\mu }({\bf P},t)\,e^{-i\omega _{\mu
}t}\,e^{-iE_{P}t/\hbar }  \label{13}
\end{equation}
and density matrix elements in momentum space are given by 
\begin{equation}
\rho _{\mu \nu }({\bf P},{\bf P}^{\prime },t)=a_{\mu }({\bf P},t)a_{\nu
}^{\ast }({\bf P}^{\prime },t)=A_{\mu }({\bf P},t)A_{\nu }^{\ast }({\bf P}%
^{\prime },t)e^{-iE_{PP^{\prime }}t/\hbar }e^{-i\omega _{\mu \nu }t}
\label{14}
\end{equation}
where $\omega _{\mu \nu }=\omega _{\mu }-\omega _{\nu }$ and $E_{PP^{\prime
}}=E_{P}-E_{P^{\prime }}.$ When the wave function (\ref{11},\ref{12}) is
substituted into Schr\"{o}dinger's equation with the Hamiltonian (\ref{3}),
one finds that the state amplitudes $A_{\mu }({\bf P},t)$ evolve as 
\begin{mathletters}
\label{15}
\begin{eqnarray}
\dot{A}_{1}({\bf P},t) &=&-i\sum_{\mu =1,2}\chi _{\mu }^{\ast }(t)\,\exp
[i\Delta _{\mu }t-i\omega _{k_{\mu }}t-i{\bf k}_{\mu }\cdot {\bf P}%
t/M]\,A_{2}({\bf P+}\hbar {\bf k}_{\mu },t)  \label{15a} \\
\dot{A}_{2}({\bf P},t) &=&-i\sum_{\mu =1,2}\chi _{\mu }(t)\,\exp [-i\Delta
_{\mu }t-i\omega _{k_{\mu }}t+i{\bf k}_{\mu }\cdot {\bf P}t/M]\,A_{1}({\bf P-%
}\hbar {\bf k}_{\mu },t),  \label{15b}
\end{eqnarray}
where 
\end{mathletters}
\begin{equation}
\Delta _{\mu }=\Omega _{\mu }-\omega  \label{16}
\end{equation}
is an atom-field detuning and 
\begin{equation}
\omega _{k_{\mu }}=\hbar k_{\mu }^{2}/2M  \label{17}
\end{equation}
is a frequency associated with atomic recoil. In terms of these state
amplitudes, the density matrix element $\rho _{21}({\bf R},t)=A_{2}({\bf R}%
,t)A_{1}^{\ast }({\bf R},t)$ can be written as 
\begin{equation}
\rho _{21}({\bf R},t)=\frac{1}{\left( 2\pi \hbar \right) ^{3/2}}\int d{\bf P}%
d{\bf P}^{\prime }A_{2}({\bf P},t)A_{1}^{\ast }({\bf P}^{\prime
},t)\,e^{-i\omega t}e^{i({\bf P}-{\bf P}^{\prime }){\bf \cdot R}/\hbar
}\,e^{-iE_{PP^{\prime }}t/\hbar }.  \label{18}
\end{equation}
Together, Eqs. (\ref{9}), (\ref{6}), (\ref{18}) and (\ref{15}) form a closed
set of equations which can be solved numerically to obtain the time
evolution of the probe field.

To simplify the analysis of both the RIR and CARL, it is assumed that 
\begin{equation}
\left| \Delta _{2}\right| \approx \left| \Delta _{1}\right| \equiv \left|
\Delta \right| \gg k_{\mu }u,\left| \chi _{\mu }\right| ,\gamma _{2};\qquad
\left| \delta \right| \ll \left| \Delta \right|  \label{19}
\end{equation}
where $\gamma _{2}$ is the excited state decay rate and $u$ is the most
probable atomic speed. If inequalities (\ref{19}) hold, both the RIR and
CARL can be interpreted in terms of {\em stimulated }processes involving the
pump and probe fields - spontaneous emission plays a negligible role. Both
the pump and probe fields are turned on and brought adiabatically to their
''initial'' values at $t=0$ in a time long compared with $\left| \Delta
\right| ^{-1}$, but small compared with all other evolution times in the
problem. Thus, at $t=0$, the values for the Rabi frequencies are 
\begin{equation}
\chi _{1}(0)=\chi _{1};\qquad \chi _{2}(0)=\chi _{2}(0),  \label{20}
\end{equation}
while the density matrix for the atoms is taken as 
\begin{equation}
\varrho _{\mu \nu }({\bf P},{\bf P}^{\prime };0)=\overline{a_{\mu }({\bf P}%
,0)a_{\nu }^{\ast }({\bf P}^{\prime },0)}=\overline{A_{\mu }({\bf P}%
,0)A_{\nu }^{\ast }({\bf P}^{\prime },0)}=(2\pi \hbar )^{3}V^{-1}W({\bf P}%
)\delta ({\bf P}-{\bf P}^{\prime })\delta _{\mu \nu }\delta _{\mu 1},
\label{21}
\end{equation}
where $V$ is the sample volume, $W({\bf P})$ is the initial momentum
distribution, and the bar indicate an ensemble average \cite{delta(0)}. This
initial density matrix corresponds to a Wigner function $W({\bf R},{\bf P}%
;0)=W({\bf P})/V$, which is the product of the momentum distribution times a
uniform spatial density (recall that this is a single particle density). The
calculation can still be carried out using an amplitude picture. The initial
condition (\ref{21}) is invoked once the density matrix element $\varrho
_{\mu \nu }({\bf R},t)$ has been evaluated.

When inequalities (\ref{19}) hold, it is possible to solve Eq.(\ref{15b})
adiabatically for the upper state amplitude in terms of the lower state
amplitude. Explicitly, one finds 
\begin{equation}
A_{2}({\bf P},t)=\sum_{\mu =1,2}[\chi _{\mu }(t)/\Delta ]\,\exp [-i\Delta
_{\mu }t-i\omega _{k}t+i{\bf k}_{\mu }\cdot {\bf P}t/M]\,A_{1}({\bf P-}\hbar 
{\bf k}_{\mu },t),  \label{22}
\end{equation}
where $\omega _{k}\equiv \omega _{k_{1}}$ and terms of order $\left| \left(
\Delta _{1}-\Delta _{2}\right) /\Delta _{1}\right| $ and $\left| (\omega
_{k_{1}}-\omega _{k_{2}})/\omega _{k_{1}}\right| $ have been ignored.
Substituting this expression back into Eq. (\ref{15a}), one obtains 
\begin{equation}
dA_{1}({\bf P},t)/dt=-i\sum_{\mu ,\nu =1,2}\frac{\chi _{\mu }(t)\chi _{\nu
}^{\ast }(t)}{\Delta }\,\exp [i\Delta _{\nu \mu }t-i\omega _{k_{\nu \mu }}t-i%
{\bf k}_{\nu \mu }\cdot pt/M]\,A_{1}({\bf P+}\hbar {\bf k}_{\nu \mu },t),
\label{23}
\end{equation}
where 
\begin{equation}
\Delta _{\nu \mu }=\Delta _{\nu }-\Delta _{\mu };\qquad {\bf k}_{\nu \mu }=%
{\bf k}_{\nu }-{\bf k}_{\mu }.  \label{24}
\end{equation}
Since ${\bf k}_{\nu \mu }$ can take on the values [0,$\pm {\bf q=}\pm ({\bf k%
}_{1}-{\bf k}_{2})$] only, it is clear that a solution to this equation can
be written as 
\begin{equation}
A_{1}({\bf P},t)=\sum_{n=-\infty }^{\infty }S_{n}({\bf P},t)A_{1}({\bf P-}%
n\hbar {\bf q},0).  \label{25}
\end{equation}
Combining Eqs. (\ref{26}), (\ref{18}), (\ref{25}) and (\ref{21}), one finds
that the density matrix element $\rho _{21}({\bf R},t)$ is given by 
\begin{eqnarray}
\rho _{21}({\bf R},t) &=&\sum_{\mu =1,2}\sum_{n,n^{\prime }=-\infty
}^{\infty }\frac{\chi _{\mu }(t)}{V\Delta }e^{i({\bf k}_{\mu }\cdot {\bf R}%
-\Omega _{\mu }t)}e^{i(n-n^{\prime }){\bf q}\cdot {\bf R}}e^{-i(n^{2}-n^{%
\prime 2})\omega _{q}t}  \nonumber \\
&&\times \int d{\bf P}W({\bf P})e^{-i(n-n^{\prime }){\bf P\cdot q}t/M}S_{n}(%
{\bf P}+n\hbar {\bf q},t)S_{n}^{\ast }({\bf P}+n^{\prime }\hbar {\bf q},t).
\label{26}
\end{eqnarray}

Equation (\ref{26}) proves that the general form for $\rho _{21}({\bf R},t)$
is correctly given by Eq. (\ref{6}). Extracting the coefficient of $e^{i(%
{\bf k}_{2}\cdot {\bf R}-\Omega _{2}t)}$ provides us with the value of $%
\tilde{\rho}_{21}(t;2)$ needed in Eq. (\ref{9}). Two types of terms in the
summation vary as $e^{i({\bf k}_{2}\cdot {\bf R}-\Omega _{2}t)}$, those with 
$\mu =2$ and $n=n^{\prime }$ and those with $\mu =1$ and $n^{\prime }=n+1.$
Denoting the $n=n^{\prime }$ term as $\tilde{\rho}_{21}(t;a)$, the $%
n^{\prime }=n+1$ as $\tilde{\rho}_{21}(t;b)$, and using the normalization
condition \cite{delta(0)} 
\begin{equation}
1=\sum_{\mu =1,2}\int d{\bf R\,}\left| A_{\mu }({\bf R},t)\right|
^{2}\approx \int d{\bf R\,}\left| A_{1}({\bf R},t)\right|
^{2}=\sum_{n=-\infty }^{\infty }\int d{\bf P}W({\bf P})\left| S_{n}({\bf P}%
+n\hbar {\bf q},t)\right| ^{2}  \label{27}
\end{equation}
one finds 
\begin{equation}
\tilde{\rho}_{21}(t;2)=\tilde{\rho}_{21}(t;a)+\tilde{\rho}_{21}(t;b),
\label{28}
\end{equation}
with 
\begin{mathletters}
\label{29}
\begin{eqnarray}
\tilde{\rho}_{21}(t;a) &=&\frac{\chi _{2}(t)}{V\Delta }\sum_{n=-\infty
}^{\infty }\int d{\bf P}W({\bf P})\left| S_{n}({\bf P}+n\hbar {\bf q}%
,t)\right| ^{2}\approx \frac{\chi _{2}(t)}{V\Delta };  \label{29a} \\
\tilde{\rho}_{21}(t;b) &=&\frac{\chi _{1}e^{i\delta t}}{V\Delta }%
\sum_{n=-\infty }^{\infty }e^{i(2n+1)\omega _{q}t}\int d{\bf P}W({\bf P}%
)e^{-i{\bf P\cdot q}t/M}S_{n}({\bf P}+n\hbar {\bf q},t)S_{n+1}^{\ast }[{\bf P%
}+(n+1)\hbar {\bf q},t].  \label{29b}
\end{eqnarray}
The first term represents linear dispersion of the medium. The second term,
which is at the heart of the RIR and CARL, represents a scattering of the
pump field off the atomic density distribution created by both the pump and
probe fields. We now have all the ingredients necessary to derive the RIR
and CARL equations.

\section{CARL and RIR}

In the large detuning limit, both the RIR and CARL equations are most easily
derived using an effective Hamiltonian for ground state atoms. An effective
Hamiltonian of the form 
\end{mathletters}
\begin{equation}
H_{eff}=\frac{P^{2}}{2M}+\frac{\hbar }{\Delta }\left( \left| \chi
_{1}\right| ^{2}+\left| \chi _{2}(t)\right| ^{2}\right) +\frac{\hbar }{%
\Delta }\left[ \chi _{1}\chi _{2}^{\ast }(t)\,e^{i({\bf q}\cdot {\bf R}%
+\delta t)}+\chi _{1}^{\ast }\chi _{2}(t)\,e^{-i({\bf q}\cdot {\bf R}+\delta
t)}\right]  \label{30}
\end{equation}
leads to the evolution equation (\ref{23}) for the ground state amplitude $%
A_{1}({\bf P},t).$ The second term in the Hamiltonian is the spatially
homogeneous light shift potential and is not of interest here. The third
term is the potential formed by the pump and probe fields which gives rise
to RIR and CARL. The depth of the potential, $U=\hbar \chi _{1}\chi
_{2}^{\ast }(t)/\left| \Delta \right| $, is a critical parameter$.$ If the
energy spread of the atoms $\delta {\cal E}$ is much larger than $U$, any
''bunching'' or spatial modulation of the atomic density will be negligibly
small, independent of the detuning $\delta $. This is generally the case at
thermal energies, making it all but impossible to observe RIR and CARL using
atoms in cells at or above the Doppler limit of laser cooling (see
discussion in Sec. V). Assuming that $\delta {\cal E}<U$, there are still
two limiting cases to consider$.$ If $\delta {\cal E}$ is much greater than
the level spacings in the potential $U$, then the motion of the atoms can be
treated classically. By approximating the potential as that of a harmonic
oscillator and assuming that the spread $\Delta P$ is of order $\hbar q$,
one can estimate that the validity condition for the motion to be treated
classically is $\left| \chi _{1}\chi _{2}/\Delta \omega _{q}\right| \gg 1$.
This is the limit considered by Bonifacio and coworkers \cite{bon} and will
be referred to as CARLSC, with the ''SC'' denoting the semiclassical limit.
A theory of CARL in which the atomic motion is fully quantized has been
given recently by Moore and Meystre \cite{meystre}. It will be referred to
as CARLQ, when it is necessary to distinguish between CARLSC and CARLQ. The
theory of RIR has been developed using a quantized description of the
center-of-mass motion; consequently, the RIR and CARLQ theories should
produce identical results, and the RIR and CARLSC theories should produce
the same results when $\left| \chi _{1}\chi _{2}/\Delta \omega _{q}\right|
\gg 1$. It should be noted that, although the atomic motion in the {\em %
optical potential} is treated classically in CARLSC, the gain in CARLSC
results from recoil-induced processes. This point is discussed in more
detail in Sec. V.

In calculating expectation values of operators that are diagonal in the
internal states, one can neglect contributions from the excited state since
the excited state population is assumed to be negligibly small. The CARL
equations are derived using the Heisenberg equations of motion while the RIR
are derived using a density matrix approach. I now show that these methods
yield equivalent results, as they must.

\subsection{CARL}

To make connection with CARL, one defines a Heisenberg operator, ''bunching
parameter,'' $B(t)=e^{i{\bf q}\cdot {\bf R}(t)}$. Using Eqs. (\ref{14}), (%
\ref{25}), (\ref{21}), and (\ref{29b}), one can write the average value of
the bunching parameter as \cite{delta(0)} 
\begin{mathletters}
\label{31}
\begin{eqnarray}
\left\langle B(t)\right\rangle  &=&\left\langle e^{i{\bf q}\cdot {\bf R(}%
t)}\right\rangle =\int d{\bf P}d{\bf P}^{\prime }{\bf \,}\rho _{11}({\bf P},%
{\bf P}^{\prime };t){\bf \,}\left\langle {\bf P}^{\prime }\right| e^{i{\bf q}%
\cdot {\bf R}}\left| {\bf P}\right\rangle   \label{31a} \\
&=&\frac{1}{(2\pi \hbar )^{3}}\int d{\bf R}d{\bf P}d{\bf P}^{\prime }%
\overline{{\bf \,}A_{1}({\bf P},t)A_{1}^{\ast }({\bf P}^{\prime },t){\bf \,}}%
e^{i{\bf q}\cdot {\bf R}}e^{i({\bf P}-{\bf P}^{\prime }){\bf \cdot R}/\hbar
}\,e^{-iE_{PP^{\prime }}t/\hbar }  \label{31b} \\
&=&\int d{\bf P\,}\overline{A_{1}({\bf P},t)A_{1}^{\ast }({\bf P}+\hbar {\bf %
q},t)}\,e^{-iE_{P,\left| {\bf P}+\hbar {\bf q}\right| }t/\hbar }  \label{31c}
\\
&=&\sum_{n,n^{\prime }=-\infty }^{\infty }\int d{\bf P\,}S_{n}({\bf P}%
,t)S_{n^{\prime }}^{\ast }({\bf P}^{\prime }+\hbar {\bf q},t)\,\rho _{11}[%
{\bf P}-n\hbar {\bf q},{\bf P}^{\prime }-(n^{\prime }-1)\hbar {\bf q}%
;0]\,e^{-iE_{P,\left| {\bf P}+\hbar {\bf q}\right| }t/\hbar } \\
&=&V^{-1}\sum_{n=-\infty }^{\infty }e^{i(2n+1)\omega _{q}t}\int d{\bf P\,}W(%
{\bf P}){\bf \,}e^{-i{\bf P\cdot q}t/M}S_{n}({\bf P}+n\hbar {\bf q}%
,t)S_{n+1}^{\ast }[{\bf P}+(n+1)\hbar {\bf q},t]  \label{31e} \\
&=&\frac{\Delta Ve^{i\delta t}}{\chi _{1}}\tilde{\rho}_{21}(t;b),
\label{31f}
\end{eqnarray}
where the brackets denote a quantum-mechanical average and the bar an
ensemble average. Therefore, by combining Eqs. (\ref{9}), (\ref{28}), (\ref
{29}), and (\ref{31f}), one finds 
\end{mathletters}
\begin{equation}
\frac{d\chi _{2}(t)}{dt}=-i\frac{nd^{2}\Omega _{2}}{2\hbar \epsilon _{0}}%
\left\{ \frac{\chi _{2}(t)}{\Delta }+\frac{\chi _{1}e^{i\delta t}}{\Delta }%
\left\langle B(t)\right\rangle \right\} ,  \label{32}
\end{equation}
where $n=N/V$ is the atomic density. This is one of the basic CARL equations
in the limit of large $\Delta $ \cite{bon}$.$ The remaining CARL equations,
obtained from the Heisenberg equations of motion with the Hamiltonian (30)
are 
\begin{mathletters}
\label{33}
\begin{eqnarray}
dB/dt &=&i\left( \frac{{\bf q\cdot P}(t)}{M}-\omega _{q}\right) B(t);
\label{33a} \\
d{\bf P}/dt &=&-i\frac{\hbar {\bf q}}{\Delta }[\chi _{1}\chi _{2}^{\ast
}(t)e^{i\delta t}B(t)-\chi _{2}(t)\chi _{1}^{\ast }e^{-i\delta t}B^{\dagger
}(t)].  \label{33b}
\end{eqnarray}
These equations do not form a closed set since the equation for $%
\left\langle dB(t)/dt\right\rangle $ involves $\left\langle {\bf P}%
(t)B(t)\right\rangle $. One must generate a series of equations for these
higher moments or use some approximation techniques (such as assuming
classical motion in the effective potential) to obtain a solution to the
equations. The linear dispersion parameter $\alpha =\frac{nd^{2}\Omega _{2}}{%
2\hbar \epsilon _{0}\Delta }$ results in a shift of the probe field
frequency. By setting $\chi _{2}(t)=\tilde{\chi}_{2}(t)e^{i\delta t}$ in
Eqs. (\ref{32},\ref{33}), it is easy to see that the detuning $\delta $
enters only in the combination $\delta +\alpha .$ Thus, one can redefine the
detuning to include the dispersion shift and replace Eq. (\ref{32}) by 
\end{mathletters}
\begin{equation}
\frac{d\chi _{2}(t)}{dt}=-i\frac{nd^{2}\Omega _{2}\chi _{1}e^{i\delta t}}{%
2\hbar \epsilon _{0}\Delta }\left\langle B(t)\right\rangle .  \label{33m}
\end{equation}

To examine the small signal gain regime, one can solve Eqs. (\ref{33}) in
perturbation theory. To zeroth order in the fields, 
\begin{mathletters}
\label{34}
\begin{eqnarray}
{\bf P}(t) &=&{\bf P(}0)\equiv {\bf P;}  \label{34a} \\
B^{(0)}(t) &=&\exp \left\{ i\left( \frac{{\bf q\cdot P}}{M}-\omega
_{q}\right) t\right\} B(0)=B(0)\exp \left\{ i\left( \frac{{\bf q\cdot P}}{M}%
+\omega _{q}\right) t\right\} ;  \label{34b} \\
\left[ B^{(0)}(t)\right] ^{\dagger } &=&\exp \left\{ -i\left( \frac{{\bf %
q\cdot P}}{M}+\omega _{q}\right) t\right\} B^{\dagger }(0)=B^{\dagger
}(0)\exp \left\{ i\left( \frac{{\bf q\cdot P}}{M}+\omega _{q}\right)
t\right\} ,  \label{34c}
\end{eqnarray}
where the last equalities in Eqs. (\ref{34b},\ref{34c}) follow from the
commutation properties of ${\bf P}$ and $B(0)=e^{i{\bf q}\cdot {\bf R(}%
0)}\equiv e^{i{\bf q}\cdot {\bf R}}.$ Note that ${\bf P}={\bf P}(0)$ and $%
{\bf R}={\bf R}(0)$ are normal Schr\"{o}dinger operators. Substituting Eqs. (%
\ref{34b},\ref{34c}) into (\ref{33b}), one finds, to first order in $\left|
\chi _{1}\chi _{2}\right| $, 
\end{mathletters}
\begin{equation}
{\bf P}^{(1)}(t)=-i\frac{\hbar {\bf q}}{\Delta }\int_{0}^{t}dt^{\prime
}\left( 
\begin{array}{c}
\chi _{1}\chi _{2}^{\ast }(t^{\prime })\,e^{i\left( \frac{{\bf q\cdot P}}{M}%
+\delta -\omega _{q}\right) t^{\prime }}B(0) \\ 
-\chi _{2}(t^{\prime })\chi _{1}^{\ast }\,e^{-i\left( \frac{{\bf q\cdot P}}{M%
}+\delta +\omega _{q}\right) t^{\prime }}B^{\dagger }(0)
\end{array}
\right) ,  \label{35}
\end{equation}
and, when this result is substituted in Eq. (\ref{33a}), one obtains 
\begin{mathletters}
\label{36}
\begin{eqnarray}
B^{(1)}(t) &=&\frac{2\omega _{q}e^{-i\delta t}}{\Delta }\int_{0}^{t}dt^{%
\prime }\exp ^{i\left( \frac{{\bf q\cdot P}}{M}+\delta -\omega _{q}\right)
(t-t^{\prime })}  \nonumber \\
&&\times \int_{0}^{t^{\prime }}dt^{\prime \prime }\left( 
\begin{array}{c}
\chi _{1}\chi _{2}^{\ast }(t^{\prime \prime })\,e^{i\left( \frac{{\bf q\cdot
P}}{M}+\delta -\omega _{q}\right) t^{\prime \prime }}B(0)e^{i\left( \frac{%
{\bf q\cdot P}}{M}+\delta -\omega _{q}\right) t^{\prime }}B(0) \\ 
-\chi _{2}(t^{\prime \prime })\chi _{1}^{\ast }\,e^{-i\left( \frac{{\bf %
q\cdot P}}{M}+\delta +\omega _{q}\right) t^{\prime \prime }}B^{\dagger
}(0)e^{i\left( \frac{{\bf q\cdot P}}{M}+\delta -\omega _{q}\right) t^{\prime
}}B(0)
\end{array}
\right)  \label{36a} \\
&=&\frac{2\omega _{q}}{\Delta }\int_{0}^{t}dt^{\prime }\exp ^{i\left( \frac{%
{\bf q\cdot P}}{M}-\omega _{q}\right) (t-t^{\prime })}  \nonumber \\
&&\times \int_{0}^{t^{\prime }}dt^{\prime \prime }\left( 
\begin{array}{c}
\chi _{1}\chi _{2}^{\ast }(t^{\prime \prime })\exp ^{i\left( \frac{{\bf %
q\cdot P}}{M}+\delta \right) (t^{\prime }+t^{\prime \prime })}e^{i\omega
_{q}(t^{\prime }-t^{\prime \prime })}\left[ B(0)\right] ^{2} \\ 
-\chi _{2}(t^{\prime \prime })\chi _{1}^{\ast }\exp ^{i\left( \frac{{\bf %
q\cdot P}}{M}+\delta +\omega _{q}\right) (t^{\prime }-t^{\prime \prime })}
\end{array}
\right)  \label{36b}
\end{eqnarray}
where the commutation properties of $B(0)$ [or $B^{\dagger }(0)]$ and ${\bf P%
}$ and the relationship $B^{\dagger }(0)B(0)=1$ have been used to go from (%
\ref{36a}) to (\ref{36b}). The bunching parameter $%
B(t)=B^{(0)}(t)+B^{(1)}(t) $ can now be averaged with the initial density
matrix (\ref{21}). The average of Eq. (\ref{34b}) for $B^{(0)}(t)$ vanishes
as does the first term in Eq. (\ref{36b}) for $B^{(1)}(t)$. On averaging the
remaining term in Eq.(\ref{36b}) for $B^{(1)}(t)$ with the initial density
matrix (\ref{21}), interchanging the order of integration, and carrying out
the integration over $t^{\prime },$one obtains 
\end{mathletters}
\begin{equation}
\left\langle B(t)\right\rangle =-\frac{2\chi _{1}^{\ast }e^{-i\delta t}}{%
\Delta }\int d{\bf P}\,W({\bf P})\int_{0}^{t}dt^{\prime }\chi _{2}(t^{\prime
})\exp ^{i\left( \frac{{\bf q\cdot P}}{M}+\delta \right) (t-t^{\prime
})}\sin \left[ \omega _{q}(t-t^{\prime })\right] .  \label{37}
\end{equation}
I will return to this equation after showing that an identical equation is
reached using a density matrix approach.

\subsection{RIR}

The RIR are usually calculated in the context of a density matrix approach.
From Eq. (\ref{31a}), it follows that 
\begin{equation}
\left\langle B(t)\right\rangle =\int \rho _{11}({\bf P},{\bf P}+\hbar {\bf q;%
}t)d{\bf P,}  \label{38}
\end{equation}
which gives the alternative form for Eq. (\ref{33m}), 
\begin{equation}
\frac{d\chi _{2}(t)}{dt}=-i\frac{nd^{2}\Omega _{2}\chi _{1}e^{i\delta t}}{%
2\hbar \epsilon _{0}\Delta }\int \rho _{11}({\bf P},{\bf P}+\hbar {\bf q;}t)d%
{\bf P.}  \label{39}
\end{equation}
To complete the RIR equations, one uses the Hamiltonian (\ref{30}), along
with Eqs. (\ref{11})-(\ref{14}), to obtain density matrix equations 
\begin{eqnarray}
\partial \rho _{11}({\bf P},{\bf P}^{\prime };t)/\partial t &=&-i\omega
_{PP^{\prime }}\rho _{11}({\bf P},{\bf P}^{\prime };t)  \nonumber \\
&&-i\frac{\chi _{1}^{\ast }\chi _{2}(t)e^{-i\delta t}}{\Delta }\left[ \rho
_{11}({\bf P+\hbar q},{\bf P}^{\prime };t)-\rho _{11}({\bf P},{\bf P}%
^{\prime }{\bf -\hbar q};t)\right]   \nonumber \\
&&-i\frac{\chi _{1}\chi _{2}^{\ast }(t)e^{i\delta t}}{\Delta }\left[ \rho
_{11}({\bf P-\hbar q},{\bf P}^{\prime };t)-\rho _{11}({\bf P},{\bf P}%
^{\prime }{\bf +\hbar q};t)\right] ,  \label{40}
\end{eqnarray}
where $\omega _{PP^{\prime }}=E_{PP^{\prime }}/\hbar $. It is easily
verified that equations for the quantities $d\left\langle B(t)\right\rangle
/dt=\int \dot{\rho}_{11}({\bf P},{\bf P}+\hbar {\bf q;}t)\,d{\bf P}$ and $%
d\left\langle {\bf P}(t)\right\rangle /dt=\int {\bf P}\dot{\rho}_{11}({\bf P}%
,{\bf P;}t)\,d{\bf P}$, obtained using Eq. (\ref{40}), are identical to Eqs.
(\ref{33}). Thus, the RIR density matrix equations are {\em totally
equivalent }to the corresponding operator CARL equations.

To evaluate $\left\langle B(t)\right\rangle $ in the perturbative limit
using Eq. (\ref{38}), one sets ${\bf P}^{\prime }={\bf P+\hbar q}$ in Eq. (%
\ref{40}) and replaces the density matrix elements in the right hand side of
that equation by their zeroth order values, 
\begin{equation}
\rho _{11}^{(0)}({\bf P},{\bf P}^{\prime };0)=(2\pi \hbar )^{3}V^{-1}W({\bf P%
})\delta ({\bf P}-{\bf P}^{\prime }),  \label{41}
\end{equation}
to obtain \cite{delta(0)} 
\begin{eqnarray}
\partial \rho _{11}^{(1)}({\bf P},{\bf P}+\hbar {\bf q;}t)/\partial t
&=&-i\omega _{P\left| {\bf P}+\hbar {\bf q}\right| }\rho _{11}({\bf P},{\bf P%
}+\hbar {\bf q};t)  \nonumber \\
&&-i\frac{\chi _{1}^{\ast }\chi _{2}(t)e^{-i\delta t}}{\Delta }\left[ W({\bf %
P+\hbar q)}-W({\bf P})\right] .  \label{42}
\end{eqnarray}
This equation is consistent with the RIR picture of Raman transitions
between center-of-mass momentum states differing by ${\bf \hbar q}$.
Integrating Eq. (\ref{42}) over {\bf P }and $t,$ and using Eq. (\ref{38}),
one reproduces Eq. (\ref{37}) for $\left\langle B(t)\right\rangle $. Since $%
\left\langle B(t)\right\rangle \sim 0$ as $\omega _{q}$ $\sim 0$, the probe
gain in the RIR and CARL is a recoil-induced effect.

The value of $\left\langle B(t)\right\rangle $ given by Eq. (\ref{37}),
which determines the small signal gain, depends critically on the ratio \cite
{cubeQ} 
\begin{equation}
r=\frac{qP_{0}}{M\omega _{q}}=\frac{2P_{0}}{\hbar q}  \label{43}
\end{equation}
where $P_{0}=Mu$ is the width of the momentum distribution and $u$ is the
most probable atomic speed. For $r\gg 1$ (''RIR limit''), the integrand in
Eq. (\ref{37}) is rapidly oscillating except when $t^{\prime }\approx t,$
allowing one to approximate the integral as 
\begin{mathletters}
\label{44}
\begin{eqnarray}
\left\langle B(t)\right\rangle  &\sim &-\frac{2\omega _{q}\chi _{1}^{\ast
}e^{-i\delta t}\chi _{2}(t)}{\Delta }\int d{\bf P}\,W({\bf P}%
)\int_{0}^{t}dt^{\prime }(t-t^{\prime })e^{i\left( \frac{{\bf q\cdot P}}{M}%
+\delta \right) (t-t^{\prime })}  \label{44a} \\
&=&i\frac{2\omega _{q}\chi _{1}^{\ast }e^{-i\delta t}\chi _{2}(t)}{\Delta }%
d\left\{ \int d{\bf P}\,W({\bf P})\int_{0}^{t}dt^{\prime }e^{i\left( \frac{%
{\bf q\cdot P}}{M}+\delta \right) (t-t^{\prime })}\right\} /d\delta 
\label{44b} \\
&=&i\frac{2\omega _{q}\chi _{1}^{\ast }e^{-i\delta t}\chi _{2}(t)}{\Delta }%
\left( \frac{M}{q}\right) d\left\{ \int_{-\infty }^{\infty
}dx\,W_{q}(P_{0}x)\int_{0}^{qut}dye^{i\left( yx+\frac{\delta y}{qu}\right)
}\right\} /d\delta ,  \label{44d}
\end{eqnarray}
where it has been assumed that the momentum distribution can be written as
the product of a one-dimensional, symmetric distribution, $W_{q}(P_{q}),$
and a two-dimensional distribution, $W_{\bot }(P_{\bot }),$ for momenta $%
{\bf P}_{\bot }$ transverse to the ${\bf \hat{q}}$ direction. Equation (\ref
{44}) is the general RIR result, in which the bunching parameter depends
locally on the field $\chi _{2}(t)$. By combining Eqs. (\ref{44}) and (\ref
{33m}), one sees that, in general, the exponential build up of the field is
not linear with time. However, if $qut\gg 1$ and $\delta /qu\lesssim 1$, the
expression for $\left\langle B(t)\right\rangle $ reduces to 
\end{mathletters}
\begin{eqnarray}
\left\langle B(t)\right\rangle  &=&i\frac{2\pi \omega _{q}\chi _{1}^{\ast
}e^{-i\delta t}\chi _{2}(t)}{\Delta }\left( \frac{M}{q}\right)   \nonumber \\
&&\times d\left\{ \frac{dW_{q}(P_{q}=\delta M/q)}{d\delta }+\frac{i}{\pi }%
\int_{-\infty }^{\infty }dx\,W_{q}(P_{0}x){\cal P}\left( x+\frac{\delta }{qu}%
\right) \right\} /d\delta ,  \label{150}
\end{eqnarray}
where ${\cal P}$ indicates a principal value, enabling one to combine Eqs. (%
\ref{150}) and (\ref{33m}) to obtain the small signal gain \cite{berman} 
\begin{equation}
g=\frac{nd^{2}\Omega _{2}}{2\epsilon _{0}}\frac{\pi q\left| \chi _{1}\right|
^{2}}{\Delta ^{2}}\frac{dW_{q}(\delta M/q)}{d\delta }.  \label{45}
\end{equation}
The gain depends linearly on the atomic density and the pump field
intensity. The gain is positive for $\delta <0,$ negative (absorption) for $%
\delta >0,$ and vanishes at $\delta =0$. This is a ''single particle'' gain
in that each atom contributes separately to the gain and there is no
threshold condition for gain that depends on atomic density.

The situation changes in the CARL limit $r\ll 1$. In that limit one can
replace $W({\bf P})$ in Eq. (\ref{37}) by $\delta ({\bf P})$ to obtain 
\begin{equation}
\left\langle B(t)\right\rangle =-\frac{2\chi _{1}^{\ast }e^{-i\delta t}}{%
\Delta }\int_{0}^{t}dt^{\prime }\chi _{2}(t^{\prime })\exp ^{i\delta
(t-t^{\prime })}\sin \left[ \omega _{q}(t-t^{\prime })\right] .  \label{46}
\end{equation}
The bunching parameter now depends nonlocally on the field amplitude $\chi
_{2}(t^{\prime })$; that is, it depends on the past history of $\chi _{2}(t)$%
. Equation (\ref{46}) is equivalent to the differential equation 
\begin{equation}
d^{2}\left\langle B(t)\right\rangle /dt^{2}=-\omega _{q}^{2}\left\langle
B(t)\right\rangle -\frac{2\chi _{1}^{\ast }\chi _{2}(t)\omega
_{q}e^{-i\delta t}}{\Delta },  \label{47}
\end{equation}
subject to the initial conditions $\left\langle B(0)\right\rangle
=d\left\langle B(0)\right\rangle /dt=0$. The small signal behavior is
determined by the coupled equations (\ref{47}) and (\ref{33m}). These
equations have already been analyzed by Bonifacio and coworkers for CARLSC
with neglect of the $\omega _{q}^{2}$ term in Eq. (\ref{47}) \cite{bon} and
by Moore and Meystre \cite{meystre} including this term. Gain occurs if one
of the roots of the cubic indicial equation, obtained from the coupled
equations, (\ref{47}) and (\ref{33m}), has a positive real value. In terms
of the quantity 
\begin{equation}
Q=\frac{nd^{2}\Omega _{2}\left| \chi _{1}\right| ^{2}}{\hbar \Delta
^{2}\epsilon _{0}\omega _{q}^{2}}  \label{48}
\end{equation}
the indicial equation is 
\begin{equation}
s^{3}+i\delta s^{2}+\omega _{q}^{2}s-i\omega _{q}^{2}(\omega _{q}Q-\delta
)=0,  \label{200}
\end{equation}
and the condition for gain is \cite{meystre} 
\begin{equation}
9\left( \delta /\omega _{q}\right) -\left( \delta /\omega _{q}\right) ^{3}+
\left[ 3+\left( \delta /\omega _{q}\right) ^{2}\right] ^{3/2}<27Q/2.
\label{49}
\end{equation}
If $Q\ll 1$, gain occurs in the range $\ \left| \delta /\omega _{q}+1\right|
<\sqrt{2Q}$ and the gain coefficient equals $\sqrt{2Q-\left( \delta /\omega
_{q}+1\right) ^{2}}\omega _{q}/2.$ On the other hand, for $Q\gtrsim 1$, gain
occurs for $\delta /\omega _{q}\geq 0,$ in contrast to the RIR limit$.$
Moreover in both cases, the gain depends nonlinearly on the atomic density
and there is a threshold condition for all values of $\delta /\omega
_{q}\neq -1.$ As such, CARL is a collective effect in the sense that gain
does not occur for atomic densities below a certain critical value. For $%
\delta /\omega _{q}=-1,$ which is the resonance condition for Raman
transitions between center-of-mass states having momenta ${\bf P}=0$ and $%
{\bf P}=\hbar {\bf q},$ there is gain irrespective of the value of $Q.$ For $%
Q\gg 1$ and $\left| \delta /\omega _{q}\right| \lesssim 1$, the gain varies
as $Q^{1/3}\omega _{q}\cos (\pi /6)=(\sqrt{3}/2)Q^{1/3}\omega _{q}$, and the
probe field undergoes exponential gain linear in time for $t\gtrsim
(Q^{1/3}\omega _{q})^{-1}.$

For early times, one can evaluate $\chi _{2}(t^{\prime })$ in Eq. (\ref{37})
at $t^{\prime }=0$, and combine Eq. (\ref{37}) with Eq. (\ref{33m}) to
obtain 
\begin{equation}
\chi _{2}(t)\sim \left\{ 1+iQ\omega _{q}^{2}\int d{\bf P}\,W({\bf P}%
)\int_{0}^{t}dt\int_{0}^{t}d\tau \exp ^{i\left( \frac{{\bf q\cdot P}}{M}%
+\delta \right) \tau }\sin \left( \omega _{q}\tau \right) \right\} \chi
_{2}(0).  \label{101}
\end{equation}
As long as the magnitude of the second term is much less than unity, the
build-up of the probe field depends linearly on the density and the
dependence of $d\chi _{2}/dt$ on $\chi _{2}$ is approximately local. In this
sense, the RIR limit is always valid for sufficiently small $t$ or atomic
density (since $Q$ is proportional to the density). It follows from Eq. (\ref
{101}) that the RIR limit is always valid if $Q\omega _{q}^{2}t^{2}/2\ll 1$.
If $W({\bf P})\sim \delta ({\bf P})$ and $\left| \delta \right| t,\omega
_{q}t\ll 1$, the RIR limit is valid when $Q\omega _{q}^{3}t^{3}/6\ll 1.$ The
condition $Q\omega _{q}^{3}t^{3}/6\ll 1$ is less restrictive than $Q\omega
_{q}^{2}t^{2}/2\ll 1$ for $Q\gg 1$.

\section{Degenerate Pump and Probe Fields, $\protect\delta =0$}

An interesting limiting case is one in which $W({\bf P})\sim \delta ({\bf P}%
) $ (CARL limit), and the pump and probe field frequencies are degenerate, $%
\Omega _{1}\approx \Omega _{2}\equiv \Omega $; $\delta =\Omega _{2}-\Omega
_{1}=0$ [recall that the detuning has been redefined to include the shift
arising from linear dispersion]. When $\delta =0$, the probe gain vanishes
identically in the RIR limit $r\gg 1$, but grows exponentially for CARLSC.
For CARLQ, it follows from Eq. (\ref{49}) that the probe gain is exponential
provided that 
\begin{equation}
Q>\frac{2}{3\sqrt{3}}  \label{50}
\end{equation}
This qualitative difference between the RIR and CARL limits is reason enough
to consider the $\delta =0$ case in some detail, but it is not the only
reason. The existence of exponential gain when $\delta =0$ is surprising at
first glance. It would seem that processes in which a pump photon is
absorbed and a probe photon emitted would be exactly cancelled by processes
in which a probe photon is absorbed and a pump photon emitted, owing to the
symmetry of the interaction when $W({\bf P})\sim \delta ({\bf P})$. It is
the purpose of this section to investigate the origin of exponential gain in
CARL when $\delta =0$. Calculations are carried out in a perturbative limit,
that is, to lowest order in the pump field intensity. As is shown below, the
results are also applicable to a wider range of problems.

\begin{figure}[tb!]
\centering
\begin{minipage}{16.0cm}
\epsfxsize= 16 cm \epsfysize= 16 cm \epsfbox{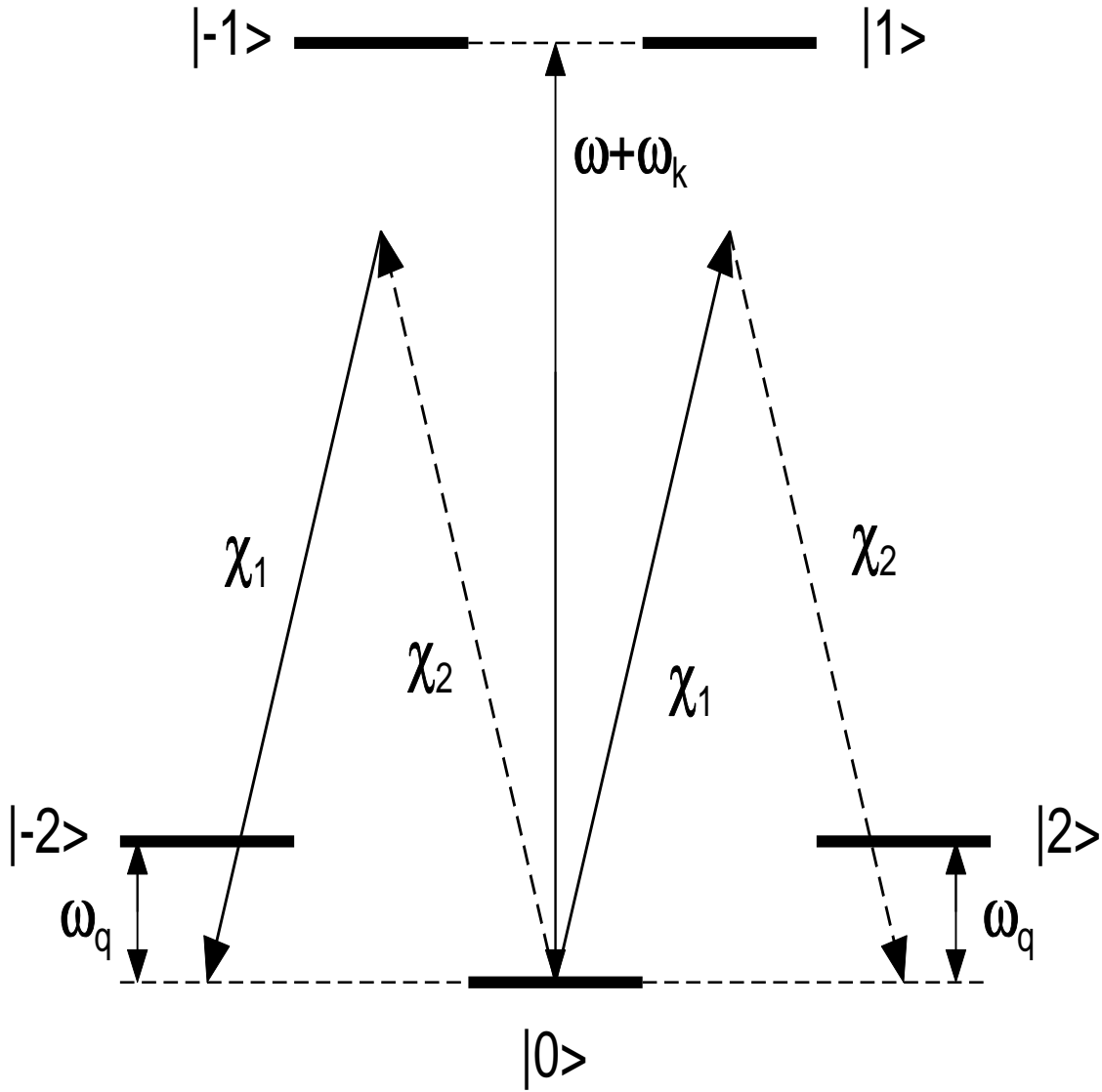}
\end{minipage}
\caption{In perturbation theory, the relevant momentum states can be
represented as an equivalent five-level system, interacting with two fields
as shown. The initial momentum distribution is taken to be a delta function
and the detuning is $\delta =0$. State $\left| 0\right\rangle $ corresponds
to $\left| g;{\bf P}=0\right\rangle ,$ state $\left| 1\right\rangle $ to $%
\left| e;{\bf P}=\hbar {\bf k}_{1}\right\rangle $, state $\left|
-1\right\rangle $ to $\left| e;{\bf P}=\hbar {\bf k}_{2}\right\rangle ,$ and
states $\left| \pm 2\right\rangle $ to $\left| g;{\bf P}=\pm 2\hbar {\bf q}%
\right\rangle $. Although derived for the recoil problem, the conclusions
reached in the text for this level scheme are applicable to a any problem
where a similar level scheme is encountered.}
\label{fig1}
\end{figure}

In perturbation theory, there are five states that enter the calculation,
starting from atoms in their ground state having ${\bf P}=0$. The relevant
states are $\left| g;{\bf P}=0\right\rangle ,$ $\left| e;{\bf P}=\hbar {\bf k%
}_{1}\right\rangle ,$ $\left| e;{\bf P}=\hbar {\bf k}_{2}\right\rangle ,$ $%
\left| g;p=\pm 2\hbar {\bf q}\right\rangle $, having energies $0$, $\hbar
(\omega +\omega _{k})$, $\hbar (\omega +\omega _{k})$, and $\hbar \omega
_{q} $, respectively (recall that $\omega _{k}\equiv \omega _{k_{1}}\approx
\omega _{k_{2}})$. It is convenient to relabel these states as 
\begin{equation}
\left| g;{\bf P}=0\right\rangle \equiv \left| 0\right\rangle ,\text{ }\left|
e;{\bf P}=\hbar {\bf k}_{1}\right\rangle \equiv \left| 1\right\rangle \text{%
, }\left| e;{\bf P}=\hbar {\bf k}_{2}\right\rangle \equiv \left|
-1\right\rangle ,\text{ }\left| g;p=\pm 2\hbar {\bf q}\right\rangle \equiv
\left| \pm 2\right\rangle \text{.}  \label{51}
\end{equation}
The energy levels associated with these states are shown in Fig. \ref{fig1}. The pump
field drives the $\left| 0\right\rangle $ to $\left| 1\right\rangle $ and $%
\left| -1\right\rangle $ to $\left| -2\right\rangle $ transitions while the
probe field drives the $\left| 0\right\rangle $ to $\left| -1\right\rangle $
and $\left| 1\right\rangle $ to $\left| 2\right\rangle $ transitions.
The Hamiltonian for the system can be obtained by expanding the Hamiltonian (\ref
{3}) in a momentum-state basis for the subspace (\ref{51}). Using the
relationship $\left\langle {\bf P}\right| e^{i{\bf k}\cdot {\bf R}}\left| 
{\bf P}^{\prime }\right\rangle =\delta ({\bf P-P}^{\prime }-\hbar {\bf k)}$,
one finds 
\begin{eqnarray}
H &=&\hbar (\omega +\omega _{k})(\left| 1\right\rangle \left\langle 1\right|
+\left| -1\right\rangle \left\langle -1\right| )+\hbar \omega _{q}(\left|
2\right\rangle \left\langle 2\right| +\left| -2\right\rangle \left\langle
-2\right| )  \nonumber \\
&&+\hbar \left\{ \chi _{1}\,e^{-i\Omega t}(\left| 1\right\rangle
\left\langle 0\right| +\left| -1\right\rangle \left\langle -2\right|
)+adjoint\right\}  \nonumber \\
&&+\hbar \left\{ \chi _{2}(t)\,e^{-i\Omega t}(\left| 1\right\rangle
\left\langle 2\right| +\left| -1\right\rangle \left\langle 0\right|
)+adjoint\right\}  \label{52}
\end{eqnarray}
Note that this level scheme could equally well describe an atom in a Stark
field, driven by circularly polarized pump and probe fields. The energy $%
\hbar \omega _{q}$ would then correspond to the Stark shifts of the $m=\pm 2$
ground state, Zeeman sublevels.

Since decay is neglected, the calculation is most conveniently carried out
using state amplitudes rather than density matrix elements. No ensemble
average is needed here since we start in an eigenstate of momentum, ${\bf P}%
=0.$ It is convenient to work in a field interaction representation in which
the state amplitudes $a_{j}({\bf P},t)$ ($j=-2,2)$ are written as 
\begin{mathletters}
\label{53}
\begin{eqnarray}
a_{0}({\bf P},t) &=&[(2\pi \hbar )^{3}/V]^{1/2}\tilde{a}_{0}(t)\,\delta (%
{\bf P})  \label{53a} \\
a_{1}({\bf P},t) &=&[(2\pi \hbar )^{3}/V]^{1/2}\tilde{a}_{1}(t)\,e^{-i\Omega
t}\,\delta ({\bf P}-\hbar {\bf k}_{1})  \label{53b} \\
a_{-1}({\bf P},t) &=&[(2\pi \hbar )^{3}/V]^{1/2}\tilde{a}_{-1}(t)\,e^{-i%
\Omega t}\,\delta ({\bf P}-\hbar {\bf k}_{2})  \label{53c} \\
a_{\pm 2}({\bf P},t) &=&[(2\pi \hbar )^{3}/V]^{1/2}\tilde{a}_{\pm
2}(t)\,\delta ({\bf P}\mp \hbar {\bf q}),  \label{53d}
\end{eqnarray}
with density matrix elements given by 
\end{mathletters}
\begin{mathletters}
\label{54}
\begin{eqnarray}
\tilde{\rho}_{\mu ,\nu }(t) &=&\tilde{a}_{\mu }(t)\tilde{a}_{\nu }^{\ast }(t)
\label{54a} \\
\rho _{\mu ,\nu }({\bf R},t) &=&\frac{1}{(2\pi \hbar )^{3}}\int d{\bf P}d%
{\bf P}^{\prime }a_{\mu }({\bf P},t)a_{\nu }^{\ast }({\bf P}^{\prime
},t)\,e^{i({\bf P}-{\bf P}^{\prime }){\bf \cdot R}/\hbar }.  \label{54b}
\end{eqnarray}
In Eq. (\ref{8a}) for the probe field evolution, one need the component of
polarization ${\bf P}_{2}$ varying as $e^{i({\bf k}_{2}{\bf \cdot R}-\Omega
t)}$. From Eqs. (\ref{53})and (\ref{54}), it follows that 
\end{mathletters}
\begin{equation}
{\bf P}_{2}({\bf R},t)=n[{\bf \wp }_{0,-1}\tilde{\rho}_{-1,0}(t)+{\bf \wp }%
_{2,1}\tilde{\rho}_{1,2}(t)]\,e^{i({\bf k}_{2}{\bf \cdot R}-\Omega t)}
\label{55}
\end{equation}
where the {\bf $\wp $}'s are dipole matrix elements. For the level scheme
corresponding to the states (\ref{51}), 
\begin{equation}
{\bf \wp }_{0,-1}={\bf \wp }_{2,1}={\bf \wp }_{eg}\equiv {\bf \wp }^{\ast }.
\label{56}
\end{equation}
Thus, the problem reduces to calculating $\tilde{\rho}_{-1,0}(t)$ and $%
\tilde{\rho}_{1,2}(t)$ to order $\left| \chi _{1}\right| ^{2}\chi _{2}$.

Before undertaking this calculation, it is useful to obtain an expression
for the time evolution of the probe field's energy density ${\cal W}_{2}$=$%
\epsilon _{0}\left| E_{2}\right| ^{2}/2$. Using Eqs. (\ref{8a}), (\ref{4}), (%
\ref{55}), and (\ref{56}), one finds 
\begin{mathletters}
\label{57}
\begin{eqnarray}
d{\cal W}_{2}/dt &=&in\Omega {\bf \wp }^{{\bf \ast }}{\bf \cdot \epsilon }%
_{2}E_{2}^{\ast }(t)(\tilde{\rho}_{-1,0}+\tilde{\rho}_{1,2})/2+c.c.
\label{57a} \\
&=&-in\hbar \Omega \chi _{2}^{\ast }(t)(\tilde{\rho}_{-1,0}+\tilde{\rho}%
_{1,2})+c.c.  \label{57b} \\
&=&n\hbar \Omega \left[ \dot{\rho}_{2,2}-\left( \dot{\rho}_{-1,-1}+\dot{\rho}%
_{-2,-2}\right) \right] .  \label{57c}
\end{eqnarray}
The last line follows from the density matrix equations of motion for the
Hamiltonian (\ref{52}) and can be given an obvious physical interpretation.
Population of state $\left| 2\right\rangle $ implies gain on the probe
field, while population in states $\left| -1\right\rangle $ or $\left|
-2\right\rangle $ implies loss for the probe field. The time rate of change
in energy density is simply the difference $\left[ \dot{\rho}_{2,2}-\left( 
\dot{\rho}_{-1,-1}+\dot{\rho}_{-2,-2}\right) \right] $ multiplied by the
product of energy, $\hbar \Omega ,$ gained or lost by the probe field in
each elementary process and the atomic density. In the adiabatic limit
considered in this paper, the excited state population is negligible and one
has 
\end{mathletters}
\begin{equation}
d{\cal W}_{2}/dt\sim n\hbar \Omega (\dot{\rho}_{2,2}-\dot{\rho}_{-2,-2}).
\label{58}
\end{equation}
One might expect that $\dot{\rho}_{2,2}=\dot{\rho}_{-2,-2}$, owing to the
symmetry of the level scheme, but we will see that this is not the case.

Assuming that the detuning of the fields $\Delta $ from the ground to
excited state atomic resonance is sufficiently large to adiabatically
eliminate the excited states, one uses the Hamiltonian (\ref{52}) to show
that the excited state amplitudes are given by 
\begin{mathletters}
\label{59}
\begin{eqnarray}
\tilde{a}_{1}(t) &\sim &\frac{\chi _{1}(t)}{\Delta }\tilde{a}_{0}(t)+\frac{%
\chi _{2}(t)}{\Delta }\tilde{a}_{2}(t);  \label{59a} \\
\tilde{a}_{-1}(t) &\sim &\frac{\chi _{2}(t)}{\Delta }\tilde{a}_{0}(t)+\frac{%
\chi _{1}(t)}{\Delta }\tilde{a}_{-2}(t),  \label{59b}
\end{eqnarray}
and that the state amplitudes $\tilde{a}_{0}(t)$, $\tilde{a}_{\pm 2}(t)$
evolve as 
\end{mathletters}
\begin{mathletters}
\label{60}
\begin{eqnarray}
d\tilde{a}_{0}(t)/dt &=&-i\chi _{1}^{\ast }\tilde{a}_{1}(t)-i\chi _{2}^{\ast
}(t)\tilde{a}_{-1}(t)  \label{60a} \\
d\tilde{a}_{2}(t)/dt &=&-i\omega _{q}\tilde{a}_{2}(t)-i\chi _{2}^{\ast }(t)%
\tilde{a}_{1}(t);  \label{60b} \\
d\tilde{a}_{-2}(t)/dt &=&-i\omega _{q}\tilde{a}_{-2}(t)-i\chi _{1}^{\ast }%
\tilde{a}_{-1}(t).  \label{60c}
\end{eqnarray}

It is a straightforward exercise to solve Eqs. (\ref{60}) in perturbation
theory to third order in the fields, starting from $\tilde{a}_{0}(0)=1$. One
finds 
\end{mathletters}
\begin{mathletters}
\label{61}
\begin{eqnarray}
\tilde{a}_{0}^{(0)}(t) &=&1  \label{61a} \\
\tilde{a}_{1}^{(1)}(t) &=&\frac{\chi _{1}}{\Delta }  \label{61b} \\
\tilde{a}_{-1}^{(1)}(t) &=&\frac{\chi _{2}(t)}{\Delta }  \label{61c} \\
\tilde{a}_{2}^{(2)}(t) &=&-i\int_{-\infty }^{t}dt^{\prime }\frac{\chi
_{2}^{\ast }(t^{\prime })\chi _{1}}{\Delta }e^{-i\omega _{q}(t-t^{\prime })}
\label{61d} \\
\tilde{a}_{-2}^{(2)}(t) &=&-i\int_{-\infty }^{t}dt^{\prime }\frac{\chi
_{1}^{\ast }\chi _{2}(t^{\prime })}{\Delta }e^{-i\omega _{q}(t-t^{\prime })}
\label{61e} \\
\tilde{a}_{0}^{(2)}(t) &=&-i\int_{-\infty }^{t}dt^{\prime }\frac{\left| \chi
_{1}\right| ^{2}+\left| \chi _{2}(t^{\prime })\right| ^{2}}{\Delta }
\label{61f} \\
\tilde{a}_{1}^{(3)}(t) &\sim &\frac{\chi _{1}}{\Delta }\tilde{a}%
_{0}^{(2)}(t)+\frac{\chi _{2}(t)}{\Delta }\tilde{a}_{2}^{(2)}(t)  \label{61g}
\\
\tilde{a}_{-1}^{(3)}(t) &\sim &\frac{\chi _{2}(t)}{\Delta }\tilde{a}%
_{0}^{(2)}(t)+\frac{\chi _{1}}{\Delta }\tilde{a}_{-2}^{(2)}(t),  \label{61h}
\end{eqnarray}
where the superscripts denote the order of the fields.

Consider, first, Eq. (\ref{58}) for the probe field intensity, which depends
on 
\end{mathletters}
\begin{equation}
d\left( \rho _{2,2}-\rho _{-2,-2}\right) /dt=d\left( \left| \tilde{a}%
_{2}^{(2)}(t)\right| ^{2}-\left| \tilde{a}_{-2}^{(2)}(t)\right| ^{2}\right)
/dt.
\end{equation}
By inspecting Eqs. (\ref{61d},\ref{61e}), one can understand the manner in
which $\left| \tilde{a}_{2}^{(2)}(t)\right| ^{2}$ can grow more rapidly than 
$\left| \tilde{a}_{-2}^{(2)}(t)\right| ^{2}$. Suppose $\chi _{2}(t)$
acquires a positive, time-dependent phase as a result of the atom-field
interaction. In this case, the quantity $\chi _{2}^{\ast }(t^{\prime
})e^{i\omega _{q}t^{\prime }}$ appearing in the integrand of Eq. (\ref{61d})
varies more slowly than the quantity $\chi _{2}(t^{\prime })e^{i\omega
_{q}t^{\prime }}$ appearing in the integrand of Eq. (\ref{61e}). As a
result, state $\left| 2\right\rangle $ population builds up more rapidly
than that of state $\left| -2\right\rangle $, leading to probe gain. In
other words, the nonlinear phase modulation of the probe field effectively
favors the $\left| 0\right\rangle $ to $\left| 2\right\rangle $ transition
over the $\left| 0\right\rangle $ to $\left| -2\right\rangle $ if the phase
is positive.

To examine the phase and amplitude build-up of the field, one can combine
Eqs. (\ref{8a}), (\ref{4}), (\ref{54})-(\ref{56}), and (\ref{61}) to obtain 
\begin{eqnarray}
\frac{d\chi _{2}}{dt} &=&-\frac{i\Omega d^{2}}{2\epsilon _{0}\hbar }n\left( 
\tilde{\rho}_{-1,0}^{(3)}+\tilde{\rho}_{1,2}^{(3)}\right)   \nonumber \\
&=&-\frac{i\Omega d^{2}}{2\epsilon _{0}\hbar }n\left[ \tilde{a}%
_{1}^{(3)}\left( \tilde{a}_{0}^{(0)}\right) ^{\ast }+\tilde{a}%
_{1}^{(1)}\left( \tilde{a}_{0}^{(2)}\right) ^{\ast }+\tilde{a}%
_{1}^{(1)}\left( \tilde{a}_{2}^{(2)}\right) ^{\ast }\right]   \nonumber \\
&=&i\frac{\Omega d^{2}}{\epsilon _{0}\hbar }n\frac{\left| \chi _{1}\right|
^{2}}{\Delta ^{2}}\int_{0}^{t}dt^{\prime }\chi _{2}(t^{\prime })\sin \left[
\omega _{q}(t-t^{\prime })\right]   \nonumber \\
&=&iQ\omega _{q}\int_{0}^{\omega _{q}t}dy^{\prime }\chi _{2}(y^{\prime
}/\omega _{q})\sin \left[ (\omega _{q}t-y^{\prime })\right] .  \label{63}
\end{eqnarray}
Of course, the integral equation (\ref{63}) is equivalent to the coupled
equations (\ref{33m},\ref{37}) with $\delta =0.$ However, this form of the
equations is convenient for obtaining the early time development of the
field and for determining the conditions under which the field undergoes
exponential gain. By solving Eq. (\ref{63}) iteratively, one finds \ 
\begin{eqnarray}
\chi _{2}(t)/\chi _{2}(0) &=&[1+iQ\omega _{q}t-(Q\omega _{q}t)^{2}/2]-iQ\sin
(\omega _{q}t)  \nonumber \\
&&-(Q^{2}/2)[4\cos (\omega _{q}t)+\omega _{q}t\sin (\omega
_{q}t)-4]+O[Q^{3}(\omega _{q}t)^{2})].  \label{64}
\end{eqnarray}
For $\omega _{q}t\ll 1$, $\chi _{2}\sim 1+i(Q\omega _{q}t)^{3}/6\sim
e^{iQ\omega _{q}t^{3}/6}$. At early times, the phase is positive, favoring
probe gain. Whether or not the field continues to grow depends on the value
of $Q.$ If $Q\ll 1,$ then $\chi _{2}$ is slowly varying compared with $\sin
(\omega _{q}t)$ and the integral in (\ref{63}) can be evaluated
asymptotically to yield 
\begin{equation}
\frac{d\chi _{2}}{dt}\sim i\omega _{q}Q[1-\cos (\omega _{q}t)]\chi _{2}\text{%
,}  \label{65}
\end{equation}
which implies that 
\begin{equation}
\chi _{2}(t)\sim e^{iQ[\omega _{q}t-\sin (\omega _{q}t)]}\chi _{2}(0)
\label{66}
\end{equation}
For $Q\ll 1$, the intensity of the probe field remains approximately
constant. In some sense, this can be viewed as the RIR limit, since the time
development of the field is approximately local, and there is neither
absorption nor gain. On the other hand, for $Q\gtrsim 1,$ the field build-up
occurs sufficiently rapidly to insure that 
\begin{eqnarray}
(\dot{\rho}_{2,2}-\dot{\rho}_{-2,-2}) &=&-2%
\mathop{\rm Re}%
[\chi _{2}^{\ast }(t)(\tilde{\rho}_{-1,0}+\tilde{\rho}_{1,2})]  \nonumber \\
&=&4%
\mathop{\rm Re}%
\left\{ i\frac{\left| \chi _{1}\right| ^{2}\chi _{2}^{\ast }(t)}{\Delta ^{2}}%
\int_{0}^{t}dy^{\prime }\chi _{2}(t^{\prime })\sin (t-t^{\prime })\right\} 
\label{67}
\end{eqnarray}
remains positive for all $t$. In this case, there is exponential gain for
the probe for times $\omega _{q}Q^{1/3}t>1$.

In summary, the probe gain that occurs for $\delta =0$ and large $\Delta $
is clearly {\em not} a single particle effect. It is more closely related to
a ''propagation'' effect in which the phase modulation of the probe field
produced by the nonlinear atom-field interaction drives the probe gain.

\section{Discussion}

It has been shown that the density matrix-RIR and Heisenberg operator-CARL
formalisms lead to equivalent equations. The RIR and CARL limits refer
simply to different regions of parameter space of these equations. For a
given experimental situation, one must determine whether one is in the RIR
limit, the CARL limit, or neither limit (as is most often the case). The
experimental implications of the RIR and CARL are discussed below, but first
I would like to discuss the distinction between the terms ''matter grating''
used in discussions of the RIR and ''atomic bunching'' used in discussions
of CARL.

The term ''matter grating'' refers to spatially modulated atom distributions
resulting from a nonlinear atom-field interaction. The term ''atomic
bunching'' refers to a redistribution or focusing of atoms in an optical
potential. For the RIR and CARL, {\em these terms are synonymous}. If recoil
effects are neglected, that is, if the center-of-mass motion is treated
classically from the outset, the {\em total atomic density} is conserved for
each velocity subclass of atoms (neglecting collisions). A homogeneous
atomic density remains homogeneous to all orders in the atom-field
interaction. Recoil effects allow for a modification of the total atomic
density. Whether one calls this ''atomic bunching'' or ''matter grating''
production is a matter of personal preference \cite{grynberg}. The key point
is that the modification of the total atomic density results entirely from
effects related to recoil on the absorption, emission, or scattering of
radiation. To lowest order in the atom-field interaction, the matter grating
or atomic bunching consists of a spatial modulation of the atomic density
having period 2$\pi /\left| {\bf k}_{1}-{\bf k}_{2}\right| $. With
increasing field strength, higher order spatial harmonics are produced,
corresponding to ''higher order matter gratings'' or ''focusing'' or
''atomic bunching.'' Of course it is possible to derive an effective
potential of the form 
\begin{equation}
H_{eff}=\left\{ \frac{\hbar }{\Delta }\left( \left| \chi _{1}\right|
^{2}+\left| \chi _{2}(t)\right| ^{2}\right) +\frac{\hbar }{\Delta }\left[
\chi _{1}\chi _{2}^{\ast }(t)\,e^{i({\bf q}\cdot {\bf R}+\delta t)}+\chi
_{1}^{\ast }\chi _{2}(t)\,e^{-i({\bf q}\cdot {\bf R}+\delta t)}\right]
\right\} \left| 1\right\rangle \left\langle 1\right|  \label{68}
\end{equation}
{\em without} quantization of the center-of-mass motion and to consider {\em %
classical} motion in this potential. However, since the effective potential
is proportional to $\hbar $, any changes in the atomic density vanish in the
classical limit. This is in contrast to bunching in the free electron laser
where the effective potential does not vanish in the classical limit.

Are there situations where matter gratings are produced by fields without
any contribution from recoil? The answer to this question is ''yes,''
provided one considers the matter gratings associated with individual
internal atomic states rather than the total atomic density. For example,for
an ensemble of {\em stationary}, closed, two-level atoms there is a spatial
modulation in the {\em population difference}$\ $between the excited and
ground states produced by the pump and probe fields. Scattering of the pump
field from this spatially modulated population difference leads to a
dispersion-shaped probe absorption profile centered at a probe-pump detuning 
$\delta =0$ in the limit of large atom-field detuning $\Delta .$ The
amplitude of the dispersion profile varies as the square of the pump field
intensity and its width is equal to the excited state decay rate \cite
{doppler}. It is also possible to have gain profiles with a width
corresponding to some effective ground state decay time if one considers
open systems in which spontaneous emission also plays a role. For example,
if the ground state consists of two hyperfine states and the fields drive
transitions between only one of these ground states and an excited state,
then, as a result of spontaneous emission, both ground state sublevel
populations can be spatially modulated (although the {\em total} atomic
density - the sum of all ground and excited state populations - remains
constant in the absence of recoil). It is possible to monitor the atomic
gratings in specific ground state hyperfine levels by using radiation that
couples only the targeted ground state sublevel to an excited state.

Finally, I would like to discuss some experimental implications of the RIR
and CARL. To observe the spectral features of the RIR and CARL, it is best
to use cold atoms in a collisionless environment. Moreover, to isolate the
effects of interest, one must use experiments which involve closed,
two-level transitions or signals that depend only on total atomic state
density. The RIR have already been observed in several experiments involving
laser-cooled atoms \cite{grynberg,gryntrans,metcalf,japanese,raizen}. The
experiments in which evidence for CARL was claimed \cite{italian,bigelow}
were all carried out under conditions (collisions, radiation trapping, large
Doppler widths) which do not favor observation of CARL. The results of these
experiments can be explained by conventional theories in which recoil is
neglected \cite{gauthier}. As was mentioned previously, the CARL limit is
essentially a subrecoil limit. Although the CARL limit has been stated as $r=%
\frac{qu}{\omega _{q}}\lesssim 1$, when $Q>1,$ a better estimate for the
CARL regime is $\frac{qu}{\omega _{q}}Q^{1/3}\lesssim 1$ \cite{cubeQ}. For
densities of order 10$^{18}$atoms/m$^{3}$, it is possible to achieve values
of $Q^{1/3}$ of order 100-1000. Thus, it may be possible to observe CARL for
atoms cooled to or somewhat above the sub-Doppler limit of laser cooling.
The linear gain coefficient in this case is of order $10^{7}$s$^{-1}$, which
must exceed any cavity loss.

Is it possible to observe the RIR in a thermal vapor using pump-probe
spectroscopy? The linear absorption rate is $g_{1}=\frac{nd^{2}\Omega _{2}}{%
\hbar \epsilon _{0}}\frac{\gamma _{2}}{\Delta ^{2}}$ which implies that the
ratio of the recoil-induced gain $g$ [Eq. (\ref{45})] to the linear
absorption rate is of order $\frac{\hbar \left| \chi _{1}\right| ^{2}/\gamma
_{2}}{E_{a}}$, where $E_{a}=mu^{2}/2$ is the kinetic energy of the atoms.
For sub-Doppler cooled atoms, this ratio can be greater than unity, but it
is small at room temperature. Still it might be possible to use modulation
techniques to isolate the RIR contribution to the probe absorption. For the
RIR signal to be larger than the background, dispersion-like contribution to
the probe absorption that varies as $2\sqrt{\pi }\left| \chi _{1}\right|
^{4}\gamma _{2}/\left( \Delta ^{5}qu\right) $, one requires that the ratio $%
\frac{\Delta ^{2}}{\left| \chi _{1}\right| ^{2}}\frac{\hbar q}{Mu}\frac{%
\Delta }{\gamma _{2}}$ be greater than unity. This can be achieved at room
temperature for sufficiently large $\Delta /\chi _{1}$.

Perhaps the best way to observe CARL would be to use an atomic beam,
transversely cooled below the recoil limit. The beam can be passed through a
cavity with some transit time $\Gamma ^{-1}.$ It is not difficult to extend
the theory to allow for a transit time effects through an effective decay
rate $\Gamma $ for ground state atoms. For subrecoil cooled atoms, Eq. (\ref
{46}) is replaced by 
\begin{equation}
\left\langle B(t)\right\rangle =-\frac{2\chi _{1}^{\ast }e^{-i\delta t}}{%
\Delta }\int_{0}^{t}dt^{\prime }\chi _{2}(t^{\prime })\exp ^{(-\Gamma
+i\delta )(t-t^{\prime })}\sin \left[ \omega _{q}(t-t^{\prime })\right] ,
\label{69}
\end{equation}
Eq. (\ref{47}) by 
\begin{equation}
d^{2}\left\langle B(t)\right\rangle /dt^{2}=-2\Gamma d\left\langle
B(t)\right\rangle /dt-\left( \omega _{q}^{2}+\Gamma ^{2}\right) \left\langle
B(t)\right\rangle -\frac{2\chi _{1}^{\ast }\chi _{2}(t)\omega
_{q}e^{-i\delta t}}{\Delta },  \label{70}
\end{equation}
and the indicial equation (\ref{200}) by 
\begin{equation}
s^{3}+(2\Gamma +i\delta )s^{2}+(\omega _{q}^{2}+\Gamma ^{2}+2i\Gamma \delta
)s-i(\omega _{q}^{3}Q-\omega _{q}^{2}\delta -\Gamma ^{2}\delta )=0.
\label{71}
\end{equation}
With the inclusion of decay, the RIR limit, in which $\left\langle
B(t)\right\rangle $ depends locally on $\chi _{2}(t),$ is $\Gamma \gg
Q^{1/3}\omega _{q}$. Thus the CARL limit occurs when $\Gamma \lesssim
Q^{1/3}\omega _{q}$. Decay tends to diminish the gain parameter when $Q\gg 1$%
, but can actually reduce the gain threshold when $Q<1$. For example, if $%
\delta =0$, the threshold condition is reduced from $Q>\frac{2}{3\sqrt{3}}$
to $Q>0$ if $\Gamma \neq 0$.

\section{Acknowledgments}

It is a pleasure to acknowledge helpful discussions with B. Dubetsky and J.
L. Cohen. This work is supported by the U. S. Office of Army Research under
Grant No. DAAG55-97-0113 and the National Science Foundation under Grants
Nos. PHY-9414020 and PHY-9800981.


\begin{references}
\bibitem{doppler}  The line width actually results from the convolution of a
Lorentzian having width equal to the excited state decay rate, $\gamma _{2},$
and a Gaussian having width $qu$, where $q$ is the magnitude of the
difference of pump and probe field wave vectors and $u$ is the most probable
atomic speed.
\bibitem{cubeQ}  A more precise statement of the CARL limit is $%
qP_{0}/M\lesssim Q^{1/3}\omega _{q}$, where $Q$ is defined by Eq. (\ref{48}%
). This condition follow from Eq. (\ref{37}), where evaluating $\chi
_{2}(t^{\prime })$ in the integrand at $t^{\prime }=t$ requires that $\chi
_{2}(t^{\prime })$ be slowly varying with respect to $\exp (i{\bf q\cdot P}%
t^{\prime }{\bf /M),}$ and the fact that $\chi _{2}(t)$ varies as $e^{\eta
Q^{1/3}\omega _{q}t}$ for $Q\gg 1$ and $Q^{1/3}\omega _{q}t>1$, where $\eta $
is a constant of order unity. Similarly, the region of simple exponential
gain for CARL occurs for times $t>(Q^{1/3}\omega _{q})^{-1}$ and, when
ground state decay is included, the field dependence is local if $\Gamma \gg
Q^{1/3}\omega _{q}$.
\end{references}
\end{document}